\documentclass[runningheads]{llncs}
\usepackage{graphicx}
\usepackage{cite}
\usepackage{mathptmx}       
\usepackage{amsmath}
\usepackage{amssymb}
\usepackage{multirow}
\usepackage{marvosym}
\usepackage{xcolor}
\begin{document}
\title{Identity Inference on Blockchain using Graph Neural Network}
\author{Jie Shen\inst{1,2} \and
Jiajun Zhou\inst{1,2} \and
Yunyi Xie\inst{1,2} \and
Shanqing Yu \inst{1,2} \textsuperscript{\Letter} \and 
Qi Xuan\inst{1,2,3} }
\authorrunning{J. Shen et al.}
%
\institute{Institute of Cyberspace Security, Zhejiang University of Technology,\\ Hangzhou 310023, China
\and College of Information Engineering, Zhejiang University of Technology, \\ Hangzhou 310023, China
\and  PCL Research Center of Networks and Communications, Peng Cheng Laboratory,\\Shenzhen 518000, China\\
\email{yushanqing@zjut.edu.cn}}

\maketitle

\begin{abstract}
The anonymity of blockchain has accelerated the growth of illegal activities and criminal behaviors on cryptocurrency platforms.
Although decentralization is one of the typical characteristics of blockchain, we urgently call for effective regulation to detect these illegal behaviors to ensure the safety and stability of user transactions.
Identity inference, which aims to make a preliminary inference about account identity, plays a significant role in blockchain security.
As a common tool, graph mining technique can effectively represent the interactive information between accounts and be used for identity inference. 
However, existing methods cannot balance scalability and end-to-end architecture, resulting high computational consumption and weak feature representation.
In this paper, we present a novel approach to analyze user's behavior from the perspective of the transaction subgraph, which
naturally transforms the identity inference task into a graph classification pattern and effectively avoids computation in large-scale graph.
Furthermore, we propose a generic end-to-end graph neural network model, named $\text{I}^2 \text{BGNN}$, which can accept subgraph as input and learn a function mapping the transaction subgraph pattern to account identity, achieving  de-anonymization. 
Extensive experiments on EOSG and ETHG datasets demonstrate that the proposed method achieve the state-of-the-art performance in identity inference.

\end{abstract}

\section{Introduction}
As a distributed database technology, blockchain achieves the function of decentralization, encryption, and tamper-proof. 
Benefiting from its anonymity, the past few years have witnessed the growing prevalence of cryptocurrencies.
As of the first quarter of 2021, there are more than 8,700 kinds of cryptocurrencies with a total market cap of 1,721 billion dollars\footnote{https://coinmarketcap.com/} \footnote{https://www.feixiaohao.com/}. 
People only need to create a pseudonymous account (synonymous with address in this paper), and they can implement transaction at almost no cost.
However, as the volume of transactions surged, the blockchain system of cryptocurrencies has also become a hotbed of illegal and criminal behavior, such as various scams\cite{vasek2015there,wu2019t,chen2018detecting} (Ponzi schemes, mining scams, scam wallets, fraudulent exchanges, etc.), money laundering\cite{bryans2014bitcoin,fanusie2018bitcoin}, abusing bot accounts\cite{huang2020understanding} and vulnerability attack\cite{maesa2017detecting}. 

As an open technique, blockchain provides public and tamper-proof transaction records, which creates the condition for data mining and analysis.
Recently, the emergence of related research has helped to analyze the transaction pattern and account behavior on the blockchain system, and most of them leverage graph modeling methods.
Such as evolution analysis of market via the on-chain transaction graph\cite{kondor2014rich,alqassem2018anti,tasca2018evolution,bai2020evolution,ferretti2020ethereum}, transaction patterns recognition via graph topology and motifs\cite{huang2017behavior,ranshous2017exchange}, detection of abnormal users or transactions via graph embedding or graph neural network\cite{yuan2020phishing,yuan2020detecting}, etc.
Among them, identity inference, which can be regarded as a de-anonymization process, is particularly important in blockchain data mining.
Generally, identity inference aims to make a preliminary inference about account identity by capturing the characteristics of the transaction pattern of the accounts.
For this task, common researches mainly concentrate on manual feature engineering including transaction features\cite{li2020identifying}, graph features\cite{pham2016anomaly} and external features\cite{ranshous2017exchange}. 
These features are mainly intuitive information, and share the same drawback like weak representation ability for classification.
Further, several methods based on random-walk\cite{wu2020phishers} and graph motif\cite{wu2021detecting} capture higher-order network features that are more representational. 
With the development of graph deep learning, graph convolution network (GCN) has attracted considerable attention and been applied in identity inference gradually, achieving outstanding results\cite{tam2019identifying,weber2019anti}. 

After reviewing the above various methods, we summarize two conflicting issues: scalability and end-to-end. 
On the one hand, real-world transaction data on blockchain systems are generally extremely huge. 
Although these methods based on feature engineering, especially manual features, show good scalability because of the independence of feature extraction, they cannot achieve end-to-end architecture.
End-to-end can reduce the reliance on expertise which is the core of feature engineering, and optimize target task in a complete form rather than multi-flows.
On the other hand, although the graph convolution network is commonly achieved via end-to-end, most of them have poor scalability.
Because the training of graph convolution network is usually performed on the whole transaction graph, where the loading and computing are not realistic.

Motivated by the subgraph perspective\cite{yuan2020phishing}, we propose a framework to reconcile the scalability and end-to-end solution for identity inference. 
Benefiting from previous work, we collect two kinds of on-chain transaction data including Ethereum and EOSIO, to infer the ``phisher'' and ``bot'' accounts, respectively. 
Firstly, we extract the transaction subgraph for each labeled accounts by a sampling mechanism. 
Through that, each account is transformed into an independent transaction subgraph. 
The sampling mechanism constrains the scale of transaction subgraph, which can effectively reduce the occupation of resources. 
Secondly, we propose an end-to-end model, to achieve \textbf{I}dentity \textbf{I}nference on \textbf{B}lockchain using \textbf{G}raph \textbf{N}eural \textbf{N}etwork (named $\text{I}^2 \text{BGNN}$).

The rest of paper is organized as follows. In Sec.~\ref{sec:related-work}, we introduce the related work about identity inference. In Sec.~\ref{sec:method}, we describe the details of our framework, including subgraph extraction and the architecture of $\text{I}^2 \text{BGNN}$. Sec.~\ref{sec:exp} presents the experiment settings and the comparison of experimental results with discussion. Finally, we conclude the paper in Sec.~\ref{sec:conclusion}.

\section{Related Work}\label{sec:related-work}
Identity inference, which aims to detect abnormal and illegal accounts, has become an effective means to monitor accounts for platform and measure transaction risks for users.
For identity inference on blockchain, related works concentrate on manual feature, graph embedding, graph neural network, and the others.

\subsubsection{Manual Feature}
Manual feature is a kind of feature engineering that relies on the experience of experts relatively. 
Normally, the more expert experience involved, the more reliable the feature vectors are. 
Lin et al.~\cite{lin2019evaluation} designed various features of transaction timestamps to express the transaction history about the accounts, and constructed a classifier against abnormal bitcoin addresses. 
Li et al.~\cite{li2020identifying} considered three kinds of features: the basic account feature, the topological feature which is related to transaction patterns, and temporal feature which is captured from the distributions of transaction timestamp. 
In addition to transaction information, Huang et al.~\cite{huang2020understanding} also considered the calling information of smart contract to expand the feature space, and finally realized the identification of bot accounts in EOSIO. 

\subsubsection{Graph Embedding}
Graph embedding aims to learn low-dimensional node representations that capture the graph structure and drive downstream graph mining task such as node classification to identify illicit accounts.
Up to now, a series of methods based on DeepWalk (DW)\cite{perozzi2014deepwalk} have been used to detect accounts. 
Yuan et al.~\cite{yuan2020detecting} used the Node2Vec algorithm which is a variant of DW to extract the potential features of the accounts and classified the phishers by Support-Vector-Machine (SVM). 
Wu et al.~\cite{wu2020phishers} redesigned the walking strategy by using transaction volume, timestamps, and multi-edges features to make their embedding framework more suitable for this task.
Subsequently, Yuan et al.~\cite{yuan2020phishing} extracted the subgraphs for each target account and embedded their transaction topology into feature vector via an embedding method named Graph2Vec~\cite{narayanan2017graph2vec}. 
Besides, they introduced the line graph~\cite{xuan2019subgraph} to further enhance the network structure embedding. 
Chen et al.~\cite{chen2020phishing} also used subgraph mechanism and got the embeddings by a graph convolution layer combining graph auto-encoder in an unsupervised way, and achieved phisher classification by LightGBM\cite{ke2017lightgbm}. 

\subsubsection{Graph Neural Network}
This part mainly about the graph neural networks with end-to-end architecture. 
In \cite{weber2019anti}, the whole transaction graph was sliced into small graphs by timestamp. 
This operation reduced the computational complexity and memory consumption which alleviate the scalability problem. 
Subsequently, graph convolution network was used for inductive learning to realize account identity inference. 
Tam et al.~\cite{tam2019identifying} used the mechanism of sampling transaction neighbors which is similar to the subgraph extraction. 
They characterized edges by embedding the temporal features from the time-series of transactions and incorporating them into the graph convolution network.
\subsubsection{Others}
Besides the aforementioned methods, there are other frameworks to achieve this identity inference. 
Phetsouvanh~\cite{phetsouvanh2018egret} proposed a graph mining technology to detect the suspicious bitcoin flow and account by analyzing the path length and confluence account of the directed subgraph. 
Zhang~\cite{zhang2020anomaly} introduced the concept of meta-path from the heterogeneous network and constructed multi-constrained meta-path based on time, attribution and topology, which is an effective way to capture behavior pattern features in a complex network.

\section{Method}\label{sec:method}
In this section, we first define the identity inference problem on blockchain, 
then present the details of subgraph extraction for constructing graph classification dataset.
Finally, we review the knowledge of using graph neural network (GNN) for learning node and graph representations, and represent the details of proposed $\text{I}^2 \text{BGCN}$ model for identity inference.

\subsection{Problem Definition}
From the perspective of graph mining, identity inference can be regarded as a node classification task. 
During node classification, the blockchain data will be modeled as a user network with million nodes, which results in unaffordable time and memory consumption for most practical algorithms.
Inspired by the core of ``neighborhood aggregation'' in graph neural network, we transform the node classification problem into a graph classification pattern in return for less time and memory consumption.

\begin{figure}
    \includegraphics[width=\textwidth]{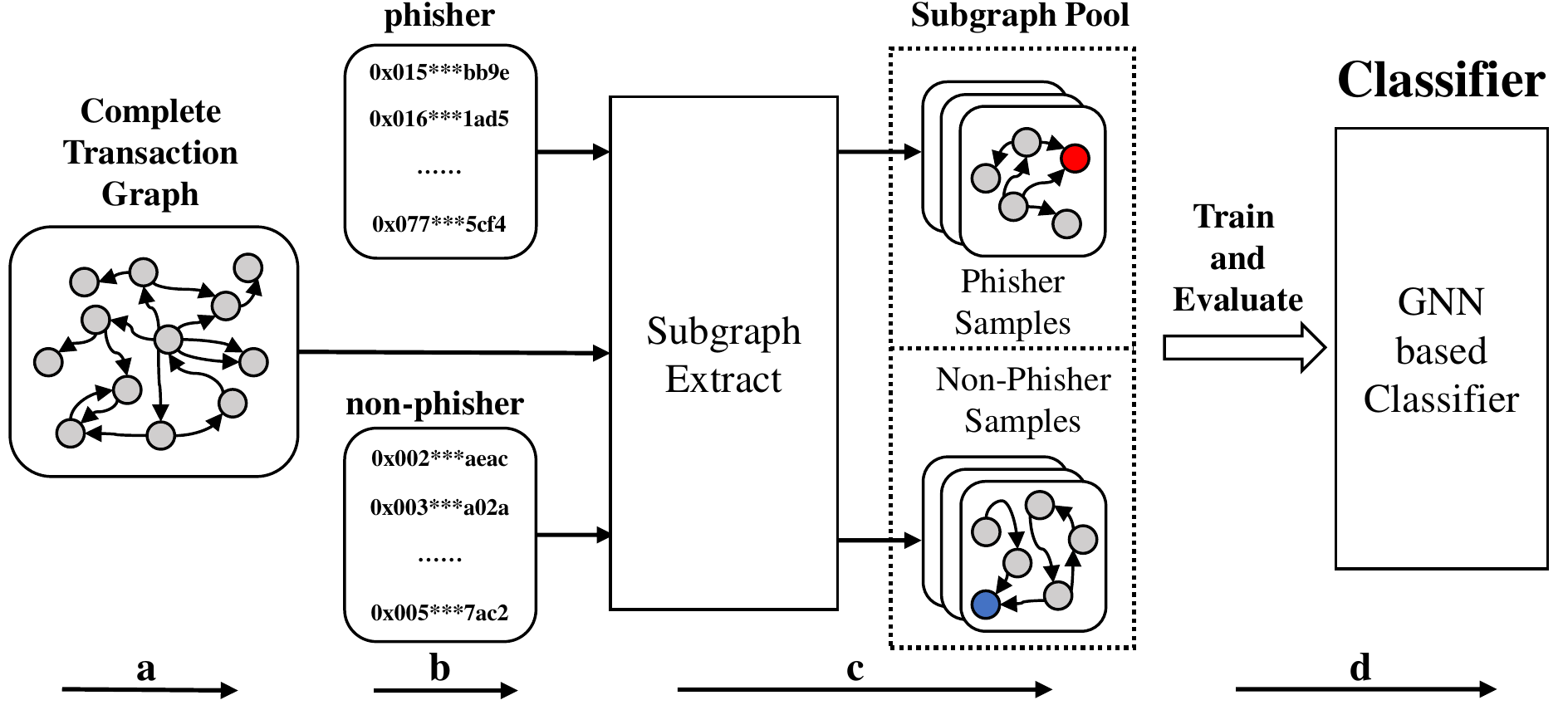}
    \caption{The schematic depiction of our framework.
             The complete workflow proceeds as follows: 
             a) modeling the transaction network; 
             b) sampling the labeled accounts;
             c) extracting the subgraphs centered on target accounts;
             d) training and evaluating using GNNs.}
    \label{fig:framework}
\end{figure}

Given a set of $n_\mathsf{A}$ labeled accounts $\mathsf{A}=\{(a_i, y_i) \mid i=1,2,\cdots, {n_\mathsf{A}}\}$, we can extract the transaction subgraph centered on each target account. 
Specifically, we extract the transaction subgraph of account $a_i$: $G_{a_i}=(V,E_v,E_t,X,y_i)$, where $V$ represents the set of accounts in this subgraph, $E_v$ and $E_t$ represent the directed edge sets that contain information about transaction volume and transaction frequency respectively, $X$ represents the calling information of smart contract, $y_i$ is the label of subgraph $G_{a_i}$.
Note that we assign the label of account $a_i$ to the transaction subgraph centered on it, and transform the node classification problem into a graph classification task: $f_\textit{nc}(a_i) \Rightarrow f_\textit{gc}(G_{a_i})$.
The final goal is to learn the transaction patterns of subgraphs and classify centered account into phishing or non-phishing via graph neural networks.
The workflow of our framework is shown in Fig.~\ref{fig:framework}, and the details of subgraph extraction and classifier design will be introduced in Subsection \ref{sec:subgraph-ext} and \ref{sec:gnn}, respectively.

\subsection{Subgraph Extraction}\label{sec:subgraph-ext}
For each account $a_i$ in $\mathsf{A}$, we check the number of its transaction partners (i.e. neighbor nodes) first.
Here, an upper limit of neighbor size, denoted as $n_u$, will be set to control the scale of transaction subgraph.
If the neighbor size of $a_i$ is less than the threshold, all neighbors and all transactions between them will be extracted.
Otherwise, we calculate and sort the total transaction volume between target account and its neighbors, and select the top-$n_u$ neighbors.
We assume that the larger the transaction volume, the higher the correlation between the two accounts.
The above extraction mechanism can also be used to sample $k$-hop transaction neighbors from the ($k-1$)-hop neighbors. 
Therefore, the scale of one-order / two-order subgraph of target account will not exceed $n_{u}$ / $(n_{u})^{2}$ normally.

The above process constructs the account (node) sets $V$, the transaction (edge) volume set $E_v$ and the transaction (edge) frequency set $E_t$.
Next, we briefly introduce two datasets which are named ETHG and EOSG and construct feature matrix $
X$ for them.

\begin{itemize}
\item \textbf{ETHG}
It is from Xblock\footnote{http://xblock.pro/} which is a blockchain data platform for academic research. 
There is an account list that contains 1660 phisher accounts and 1700 non-phisher accounts with their 2-hop transaction records in Ethereum. 
Based on this, we filter the Contract-Account (CA) which will be considered as the contract calling feature of Externally-Owned-Accounts (EOA).
And according to the span of the block where the transaction record is located, we collect all CAs from 0 to 10,000,000 blocks, filter them via the calling amount, and retain the top 14885 finally. 
After that, we construct the feature matrix of contract calling (cc) $X_\textit{cc} \in \mathbb{R}^{n\times 14885}$, and each EOA has a 14885 dimension vector to represent their calling situation about those CAs.
\item \textbf{EOSG}
It is collected by \cite{huang2020understanding}. 
They integrate and model the on-chain data of EOSIO: Enhanced Money Flow Graph (EMFG) which contains the transactions between accounts including timestamps and volume, Enhanced Account Creation Graph (EACG) which contains account creation tree data, Enhanced Contract Invocation Graph (ECIG) which contains smart contract calling data, and a list of labeled accounts which contains 229,907 normal accounts and 63863 bot-like accounts. 
Similarly, we extract the subgraph graph and contract calling features from EMFG and ECIG respectively, and construct the feature matrix of contract calling (cc) $X_\textit{cc} \in \mathbb{R}^{n\times 1213}$. 
Further, we consider the account name restriction mechanism of EOSIO and add three kinds of node labels to expand features, since that the type of neighbors can also express the transaction pattern of the account.
The three labels are the general account which consists of 12 characters, the auction account which is less than 12 characters but does not contain the character '.', and the sub-account of auction account which combines '.' with auction account name as the suffix. 
On the other hand, the neighbor extraction will stop at the system account whose name begins with 'EOSIO.'. 
Because the behavior pattern of the current center account has nothing to do with the transactions between other further accounts and system accounts. We construct the feature matrix of node label (nl) $X_\textit{nl} \in \mathbb{R}^{n\times 3}$.
\end{itemize}
In summary, the feature matrix is $X=X_\textit{cc} \in \mathbb{R}^{n\times 14885}$ for ETHG and $X=X_\textit{cc} \oplus X_\textit{nl} \in \mathbb{R}^{n\times 1216}$ for EOSG where $\oplus$ is concatenation operation.

\subsection{Graph Neural Networks} \label{sec:gnn}
By viewing the accounts interaction as graph data (i.e., transaction graph),
recent deep learning methods for graph structural data, such as graph neural network (GNNs)\cite{kipf2016semi, velickovic2018graph,gilmer2017neural}, can be utilized to learn transaction pattern representation that can be fed to downstream machine learning models for phishing account detection.
In this section, we will present the details of employing GNNs to obtain transaction pattern representation.
 
GNNs learn the representations of nodes by leveraging both the graph structure and node/edge features.
This is done by a neighborhood aggregation function that iteratively takes the representation of all neighbors together with the graph structure as input, and outputs the aggregate representation of target node.
The aggregation function can be defined using Graph Convolution layer\cite{kipf2016semi}, Graph Attention layer\cite{velickovic2018graph}, or any general message passing layer\cite{gilmer2017neural}.
Formally, a graph convolution network (GCN) model follows the following rule to aggregate the feature of neighbors:
\begin{equation}\label{eq:gcn-layer}
    H^{(l)}=\sigma (\hat{A} H^{(l-1)} W^{(l-1)}),\\
\end{equation}
where $H^{(l-1)} \in \mathbb{R}^{n\times k}$ is a matrix containing the $k$-dimensional representation of $n$
nodes in the $(l-1)$-th layer, $\sigma$ is the activation function (typically ReLU), 
$\hat{A}$ is a symmetric normalization of $A$ and can be defined as:
\begin{equation}\label{GCN-A}
    \hat{A}=\tilde{D}^{-\frac{1}{2}}\tilde{A}\tilde{D}^{-\frac{1}{2}},\ \ 
    \tilde{A} = A+I_{n},\ \ 
    \tilde{D}=\text{diag}(\sum_{j = 0}^{n}  \tilde{A}_{ij}),
\end{equation}
where $\tilde{A}$ is an $n\times n$ adjacency matrix of the graph with self connections added, $\tilde{D}$ is a degree diagonal matrix. 
After $l$ layer of computation, the node representations $H^{(l)}$ is able to capture the information within their $l$-hop neighborhoods.

Generally, GCN model is used to learn the node representations in semi-supervised node classification.
A 2-layer GCN model with softmax function can be formulated as:
\begin{equation}\label{2layer-GCN}
    Z = \text{softmax}(\hat{A}\cdot \text{ReLU}(\hat{A}XW^{(0)})W^{(1)})
\end{equation}
where $Z\in\mathbb{R}^{n\times y}$ is the prediction probability distribution and $y$ is the dimension of node labels. $W^{(0)}$ and $W^{(1)}$ are the input-to-hidden and hidden-to-output weights, respectively.
\begin{figure}
    \includegraphics[width=\textwidth]{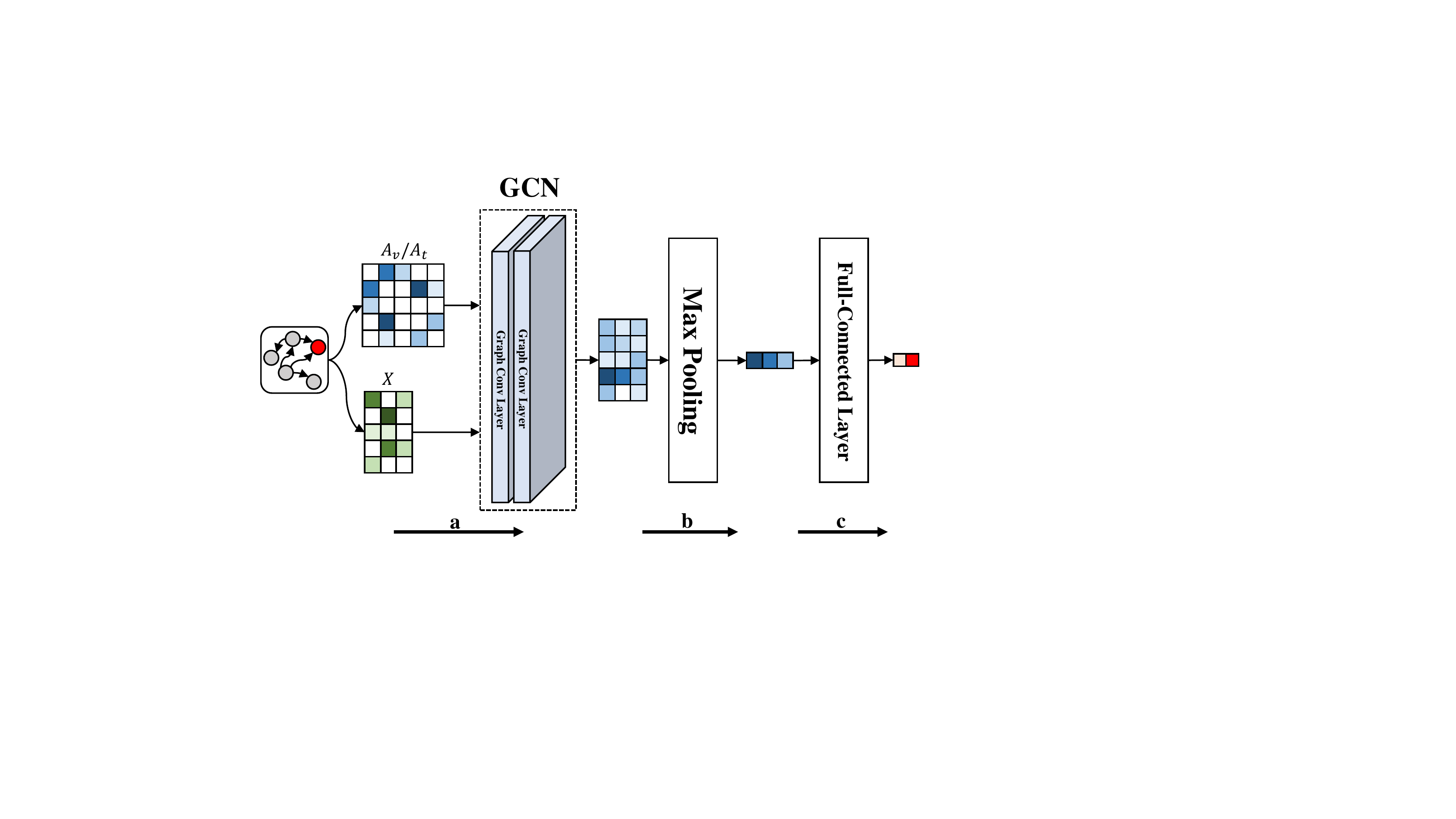}
    \caption{The architecture of $\text{I}^2 \text{BGNN}$ model. a) $A_{v}$ ($A_{t}$) and $X$ are captured from subgraph extraction and sent to graph convolution network; b) the max-pooling layer is used to compress the aggregated node representations to obtain the whole graph representation; c) the graph representation is used to predict the subgraph (account) label.}
    \label{fig:gcn}
\end{figure}

\subsection{$\text{I}^2 \text{BGNN}$}
We now present the details of proposed $\text{I}^2 \text{BGNN}$ for identity inference on blockchain.
For graph classification, the pooling operations aggregate node representations from the final iteration to obtain the whole graph’s representation.
By stacking the pooling layer and fully-connected layer after 2-layer GCN, the basic graph classification model for identity inference can be constructed as follows:
\begin{equation}\label{2layer-gcnn}
    Z = \text{softmax}(\text{MaxPooling}(ReLU(\hat{A}\cdot \text{ReLU}(\hat{A}XW^{(0)})W^{(1)}))W^{(2)}+b)
\end{equation}
Note that we use the max pooling to obtain the whole graph's representation.
The model architecture of $\text{I}^2 \text{BGNN}$ is shown in Fig.~\ref{fig:gcn}.
In a transaction subgraph, each node represents an account and each directed edge represents transaction flow that contains information about transaction volume and frequency.
For the input layer of GCN, we first initialize the node representations using their attributions in transaction subgraph. 
Specifically, the node attributions include contract calling information (cc) and distinctive node-label (nl), as mentioned in Sec.~\ref{sec:subgraph-ext}, and we initialize node representation as $H^{(0)} = X$.

\begin{table}
    \renewcommand\arraystretch{1.1}
    \centering
    \caption{Dataset properties.
             $|G|$ is the number of subgraphs in dataset,
             $Avg.|V|$ is the average number of nodes per graph,
             $Avg.|E_\textit{di}|$ is the average number of edges per directed graph,
             $Avg.|E_\textit{ud}|$ is the average number of edges per undirected graph which is transformed from corresponding directed graph,
             $|F|$ is the dimension of node features,
             $|Y|$ is the number of classes for labels.
             }\label{tb:Dataset-properties}
    \resizebox{0.8\linewidth}{!}{%
    \begin{tabular}{cccccccc} 
    \hline\hline
    \textbf{Dataset} \ &\  $|G|$ \ &\  $Avg.|V|$ \ &\  $Avg.|E_\textit{di}|$ \ &\  $Avg.|E_\textit{ud}|$ \ &\  $|F|$ \ &\  $|Y|$ \ &\  \textbf{Label bias}  \\ 
    \hline
    ETHG             & 3266           & 80                  & 239                      & 222                      & 14885                              & 2              & 0.99                 \\
    EOSG             & 2000           & 260                 & 4250                     & 3212                     & 1216                               & 2              & 1                    \\
    \hline\hline
    \end{tabular}
    }
\end{table}
\section{Experiment}\label{sec:exp}
\subsection{Dataset}
For EOSG dataset, we filter the labeled account list in term of subgraph size and obtain over 20,000 available accounts. 
Then, 1000 accounts per label are selected randomly for the follow-up experiments. 
The detailed dataset properties are given in Table~\ref{tb:Dataset-properties}.
Finally, each dataset is split into training and testing sets with a proportion of 1:1, and they will be resplit 3 times using different random seeds. 
We report the average accuracy across all trials.

\subsection{Baseline}
Since we implement identity inference with a graph classification pattern, we compare our framework with several SOTA graph classification algorithms including SF~\cite{de2018simple}, Graph2vec~\cite{narayanan2017graph2vec}, Netlsd~\cite{tsitsulin2018netlsd} and FGSD~\cite{verma2017hunt}.
The first two are graph embedding methods and the last two are graph kernel methods.
Graph2vec extends the document embedding methods to graph classification and learns a distributed representation of the whole graph via document embedding neural networks.
SF performs graph classification by spectral decomposition of the graph Laplacian, i.e., it relies on spectral features of the graph.
Netlsd performs graph classification by extracting compact graph signatures that inherit the formal properties of the Laplacian spectrum.
FGSD calculates the Moore-Penrose spectrum of the normalized laplacian and uses the histogram of the spectral features of this spectrum to represent the whole graph. 

\subsection{Experiment setting}
During subgraph extraction, the direction of edges in subgraph is determined by the transaction flow. 
However, during the experiments, we find that symmetric adjacency matrix of subgraph usually outperforms directed adjacency matrix. 
Therefore, we transform the directed adjacency matrix into a symmetric adjacency matrix by adding its transpose to itself.
Other settings for models and datasets are as follows:
\subsubsection{Method settings} For all the four baseline methods, we set the embedding dimension to 128, and use default settings for other parameters. 
Further, we implement graph classification by using the following machine learning classifiers: Support Vector Machine (SVM) with radial basis kernel, k-Nearest Neighbors classifier (KNN) and Random Forest classifier (RF). 
As for $\text{I}^2 \text{BGNN}$, we apply two layers of GCNs with output dimensions both equal to 128, and set the maximum number of eopchs to be 50, the batch size to be 30 and dropout to be 0.3.

\begin{table}
    \renewcommand\arraystretch{1}
    \centering
    \caption{Results of identity inference. The top-2 best results are highlighted in bold.}\label{tb:comparison of results}
    \resizebox{0.8\textwidth}{!}{%
    \begin{tabular}{c|c|ccc|ccc} 
    \hline\hline
    \multicolumn{2}{c|}{\multirow{3}{*}{\scriptsize{Method}}} & \multicolumn{6}{c}{\scriptsize{Dataset}}                                    \\ 
    \cline{3-8}
    \multicolumn{2}{c|}{}                        & \multicolumn{3}{c|}{\scriptsize{EOSG}}     & \multicolumn{3}{c}{\scriptsize{ETHG}}       \\ 
    \cline{3-8}
    \multicolumn{2}{c|}{}               &  ~\scriptsize{F1} & \scriptsize{Precision} & \scriptsize{Recall}~ & ~\scriptsize{F1} & \scriptsize{Precision} & \scriptsize{Recall}~  \\ 
    \hline
    \multirow{3}{*}{~\scriptsize{Graph2vec}} & ~\scriptsize{SVM}~    & ~\scriptsize{0.8223}   & \scriptsize{0.8132}    & \scriptsize{0.8317}~ & ~\scriptsize{0.6487}   & \scriptsize{0.7564}    & \scriptsize{0.5678}~  \\
                               & ~\scriptsize{KNN}~    & ~\scriptsize{0.6171}   & \scriptsize{0.9820}    & \scriptsize{0.4499}~ & ~\scriptsize{0.5705}   & \scriptsize{0.5709}    & \scriptsize{0.5701}~  \\
                               & ~\scriptsize{RF}~     & ~\scriptsize{0.7637}   & \scriptsize{0.8155}    & \scriptsize{0.7180}~ & ~\scriptsize{0.6104}   & \scriptsize{0.7355}    & \scriptsize{0.5216}~  \\ 
    \hline
    \multirow{3}{*}{~\scriptsize{SF}}        & ~\scriptsize{SVM}~    & ~\scriptsize{0.9428}   & \scriptsize{0.9222}    & \scriptsize{0.9643}~ & ~\scriptsize{0.6287}   & \scriptsize{0.6405}    & \scriptsize{0.6173}~  \\
                               & ~\scriptsize{KNN}~    & ~\scriptsize{0.9089}   & \scriptsize{0.9081}    & \scriptsize{0.9098}~ & ~\scriptsize{0.6238}   & \scriptsize{0.6401}    & \scriptsize{0.6083}~  \\
                               & ~\scriptsize{RF}~     & ~\scriptsize{0.9333}   & \scriptsize{0.9166}    & \scriptsize{0.9507}~ & ~\scriptsize{0.6908}   & \scriptsize{0.7056}    & \scriptsize{0.6766}~  \\ 
    \hline
    \multirow{3}{*}{~\scriptsize{Netlsd}}    & ~\scriptsize{SVM}~    & ~\scriptsize{0.8730}   & \scriptsize{0.8574}    & \scriptsize{0.8891}~ & ~\scriptsize{0.7067}   & \scriptsize{0.6869}    & \scriptsize{0.7276}~  \\
                               & ~\scriptsize{KNN}~    & ~\scriptsize{0.8406}   & \scriptsize{0.8417}    & \scriptsize{0.8396}~ & ~\scriptsize{0.6774}   & \scriptsize{0.6884}    & \scriptsize{0.6667}~  \\
                               & ~\scriptsize{RF}~     & ~\scriptsize{0.8845}   & \scriptsize{0.8625}    & \scriptsize{0.9077}~ & ~\scriptsize{0.6702}   & \scriptsize{0.6782}    & \scriptsize{0.6623}~  \\ 
    \hline
    \multirow{3}{*}{~\scriptsize{FGSD}}      & ~\scriptsize{SVM}~    & ~\scriptsize{0.9617}   & \scriptsize{0.9534}    & \scriptsize{0.9701}~ & ~\scriptsize{0.7206}   & \scriptsize{0.6810}    & \scriptsize{0.7650}~  \\
                               & ~\scriptsize{KNN}~    & ~\scriptsize{0.9469}   & \scriptsize{0.9404}    & \scriptsize{0.9534}~ & ~\scriptsize{0.7161}   & \scriptsize{0.6750}    & \scriptsize{0.7625}~  \\
                               & ~\scriptsize{RF}~    & ~\scriptsize{0.9578}   & \scriptsize{0.9579}    & \scriptsize{0.9578}~ & ~\scriptsize{0.7372}   & \scriptsize{0.7448}    & \scriptsize{0.7297}~  \\ 
    \hline
    \multicolumn{2}{c|}{\scriptsize{$\text{I}^2 \text{BGNN}$-v}}  & ~\textbf{\scriptsize{0.9940}}   & \textbf{\scriptsize{0.9894}}    & \textbf{\scriptsize{0.9986}}~ & ~\textbf{\scriptsize{0.8587}}   & \textbf{\scriptsize{0.8190}}    & \textbf{\scriptsize{0.9024}}~  \\
    \multicolumn{2}{c|}{\scriptsize{$\text{I}^2 \text{BGNN}$-t}}  & ~\textbf{\scriptsize{0.9950}}   & \textbf{\scriptsize{0.9917}}    & \textbf{\scriptsize{0.9983}}~ & ~\textbf{\scriptsize{0.8600}}   & \textbf{\scriptsize{0.8697}}    & \textbf{\scriptsize{0.8505}}~  \\
    \hline\hline
\end{tabular}
    }
\end{table}

\subsubsection{Metric settings} Both the datasets have two classes, so we evaluate the results of binary classification by precision, recall and F1-Score.

\subsection{Result and discussion}
\subsubsection{Inference Performance}
Table~\ref{tb:comparison of results} reports the performance comparison between $\text{I}^2 \text{BGNN}$ and baselines, from which we can observe that $\text{I}^2 \text{BGNN}$ significantly outperforms other methods across the two datasets.
Specifically, compared with baselines, our $\text{I}^2 \text{BGNN}$ achieves average improvement of 12\% / 19\% in term of F1 on EOSG / ETHG.
This may be due to the excellent expression ability of the graph convolution layer and the effectiveness of the features which are constructed by the contract calling information.
In addition, these graph embedding and kernel methods which are based on spectral analysis are significantly better than Graph2vec. Their advantages are also reflected in the efficiency of model operation in experiments.

Furthermore, we investigate the influence of neighborhood depth and data division on the experimental results under various settings.
\subsubsection{The influence of neighborhood depth}
Normally, the subgraph containing 3-hop neighbors will have a large scale, which leads to difficulties in feature learning.
To further analyze the influence of different depth for subgraph extraction, we extract 1-hop and 2-hop neighbors to construct the subgraphs for each target account. 
Table~\ref{tb:1hop-properties} shows the properties of 1-order and 2-order subgraphs for two datasets. 
And Table~\ref{tb:neighborhood depth} reports the performance comparison between 1-order and 2-order subgraphs using $\text{I}^2 \text{BGNN}$.
For EOSG, $\text{I}^2 \text{BGNN}$ with 1-order subgraph performs slightly better than that with 2-order subgraph. Actually, the 1-order transaction subgraph contains sufficient and effective characteristics of transaction behavior, while the larger scale of the 2-order subgraph leads to the redundancy of information.
As for ETHG, the situation is just the opposite, $\text{I}^2 \text{BGNN}$ with 2-order subgraph outperforms that with 1-order subgraph. 
Obviously, the 1-order subgraph contains sparse transaction information, which is not conducive to inference, while the denser interactions in the 2-order subgraph facilitate behavior analysis and identity inference.
\begin{table}
    \renewcommand\arraystretch{1.2}
    \centering
    \caption{The properties of 1-hop and 2-hop subgraphs.}
    \label{tb:1hop-properties}
    \begin{tabular}{ccccc} 
    \hline\hline
    ~\textbf{Dataset}~    & ~\textbf{Subgraph}~ & ~$Avg.|V|$~ & ~$Avg.|E_\textit{di}|$~ & ~$Avg.|E_\textit{ud}|$~  \\ 
    \hline
    \multirow{2}{*}{EOSG} &  1-order    & 17        & 65                    & 48                     \\
                          &  2-order    & 260       & 4250                  & 3212                   \\ 
    \hline
    \multirow{2}{*}{ETHG} &  1-order    & 10        & 13                    & 12                     \\
                          &  2-order    & 80        & 239                   & 222                    \\
    \hline\hline
    \end{tabular}
    \vspace{-0.8cm}
\end{table}

\begin{table}
    \renewcommand\arraystretch{1.2}
    \centering
    \caption{Results of $\text{I}^2 \text{BGNN}$ with different neighborhood depth.}\label{tb:neighborhood depth}
    \begin{tabular}{c|c|cc} 
    \hline\hline
    ~Dataset~               & ~Method~  & ~1-order~             & ~2-order~              \\ 
    \hline
    \multirow{2}{*}{EOSG} & ~$\text{I}^2 \text{BGNN}$-v~  & \textbf{0.9960} & 0.9940            \\
                          & ~$\text{I}^2 \text{BGNN}$-t~  & \textbf{0.9980} & 0.9950            \\ 
    \hline
    \multirow{2}{*}{ETHG} & ~$\text{I}^2 \text{BGNN}$-v~  & 0.8356           & \textbf{0.8587}  \\
                          & ~$\text{I}^2 \text{BGNN}$-t~  & 0.8366           & \textbf{0.8600}  \\
    \hline\hline
    \end{tabular}
    \vspace{-0.8cm}
\end{table}
\begin{figure}
    \centering
    \includegraphics[width=0.7\textwidth]{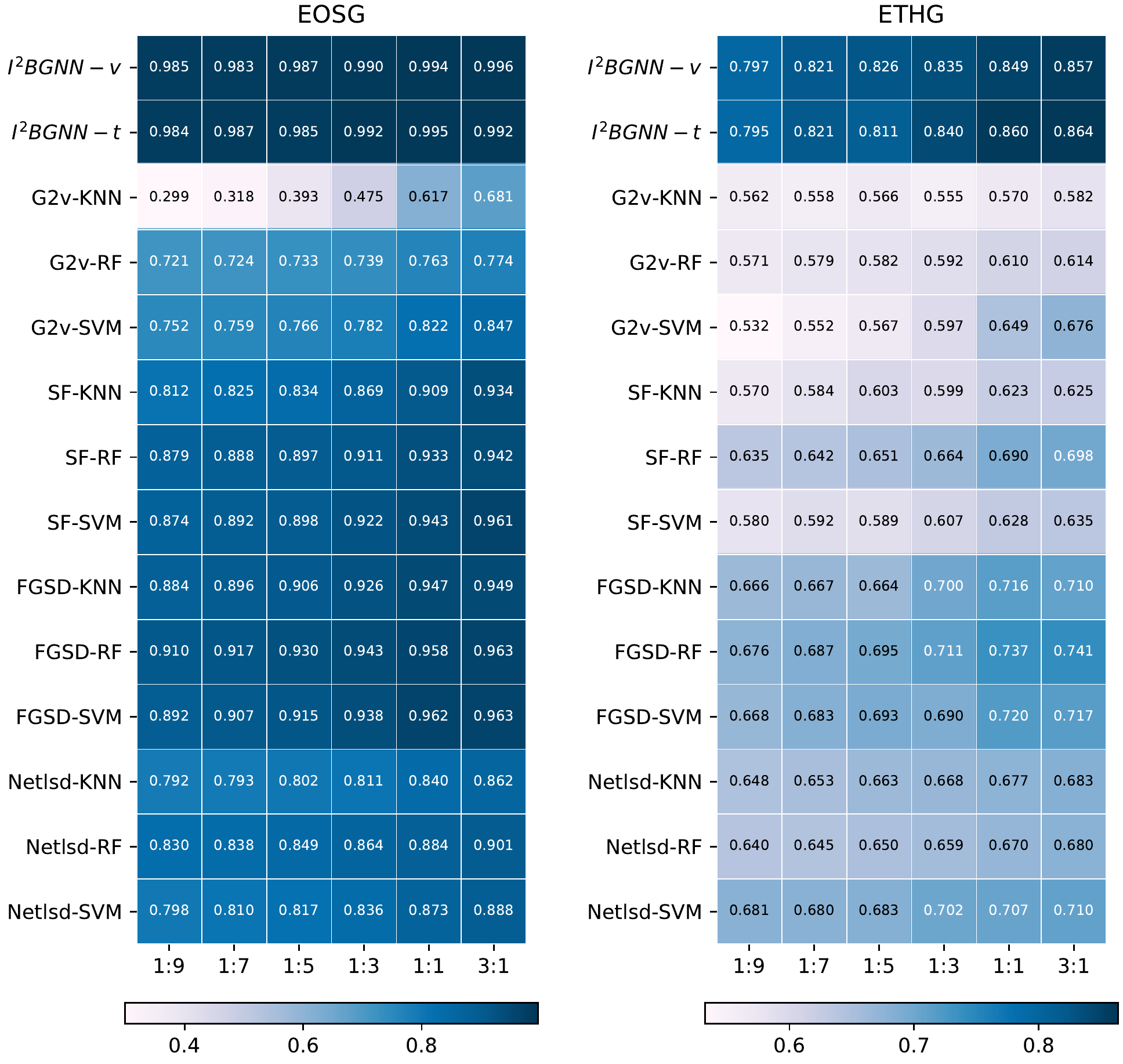}
    \caption{Experimental results of different division ratios of the dataset on EOSG(left) and ETHG(right)}
    \label{fig:datasplit}
\end{figure}
\subsubsection{The influence of data split}
Next, we analyze the sensitivity of models to different ratios of data split.
Specifically, we vary the ratio of training set to testing set in $\{\text{1:9, 1:7, 1:5, 1:3, 1:1, 3:1}\}$.
Fig.~\ref{fig:datasplit} reports the inference results (F1) of different models with various proportion of training set.
Obviously, for different ratios, our $\text{I}^2 \text{BGNN}$ holds the best performance compared with other graph classification models.
In addition, with the increase of training data, the performances of all models are naturally improved.

\section{Conclusion}\label{sec:conclusion}
Traditional graph mining methods for identity inference are stuck in a dilemma where it is difficult to integrate  scalability and end-to-end architecture into one model.
In this work, we balance scalability and end-to-end architecture in model design.
Specifically, we propose to learn the transaction subgraph centered on target account and transform the identity inference task on blockchain into graph classification pattern, resulting in a great reduction in resource consumption.
Moreover, we design an end-to-end $\text{I}^2 \text{BGNN}$ model, which is capable of learning an effective graph representation.
Finally, we conduct extensive experiments on two real blockchain datasets (EOSG and ETHG) to demonstrate the effectiveness of our proposed $\text{I}^2 \text{BGNN}$.
Experimental results show that the transaction pattern hidden in subgraph can actually reveal the account behavior, and our $\text{I}^2 \text{BGNN}$ achieves the outstanding performance in identity inference.

\subsubsection{Acknowledgements. }
The authors would like to thank all the members in the IVSN Research Group, Zhejiang University of Technology for the valuable discussions about the ideas and technical details presented in this paper. This work was partially supported by the National Key R\&D Program of China under Grant No. 2020YFB1006104, by the National Natural Science Foundation of China under Grant No. 61973273, by the Zhejiang Provincial Natural Science Foundation of China under Grant No. LR19F030001, by the Ministry of Public Security's Research Project ``Research and Demonstration Application of Key Technologies of Criminal Social Network Model'', and by the Special Scientific Research Fund of Basic Public Welfare Profession of Zhejiang Province under Grant LGF20F020016.

\bibliographystyle{unsrt}
\bibliography{ref}
\end{document}